# Three Hours a Day: Understanding Current Teen Practices of Smartphone Application Use

## Frank Bentley, Karen Church, Beverly Harrison, Kent Lyons, Matthew Rafalow

{fbentley, kchurch}@yahoo-inc.com, kent@techna.org, beverly_harrison@yahoo.com, mrafalow@uci.edu

Yahoo, Sunnyvale, CA

## Abstract

Teens are using mobile devices for an increasing number of activities. Smartphones and a variety of mobile apps for communication, entertainment, and productivity have become an integral part of their lives. This mobile phone use has evolved rapidly as technology has changed and thus studies from even 2 or 3 years ago may not reflect new patterns and practices as smartphones have become more sophisticated. In order to understand current teen's practices around smartphone use, we conducted a two-week, mixed-methods study with 14 diverse teens. Through voicemail diaries, interviews, and real-world usage data from a logging application installed on their smartphones, we developed an understanding of the types of apps used by teens, when they use these apps, and their reasons for using specific apps in particular situations. We found that the teens in our study used their smartphones for an average of almost 3 hours per day and that two-thirds of all app use involved interacting with an average of almost 10 distinct communications applications. From our study data, we highlight key implications for the design of future mobile apps or services, specifically new social and communications-related applications that allow teens to maintain desired levels of privacy and permanence on the content that they share.

## Introduction

Teens are an important demographic group representing between 17% of the total population for more developed countries (e.g., ~ 25 million in the US) to about 30% of the total population for less developed countries (World Youth Data Sheet, 2013). Thus teens represent a large consumer population with estimated spending of $117.6 billion in 2012 in the US (Statista, 2012). Perhaps more interesting, teens are often viewed as early adopters of what will later become predictive of more "mainstream" technology use patterns. According to PEW data (Madden et al., 2013), 78% of American teens have a mobile phone and 58% of teens have downloaded an application on their phone or tablet. While there are several surveys, very little has been studied about the practices teens have developed with their smartphones and applications. Also, most survey data (for example, ComScore, 2013; Nielsen, 2009 & 2014) reports on app usage from the general population, without specific focus on this unique demographic. Other reports have suggested that teens have recently become less engaged users of Facebook (Madden et al., 2013; Wortham, 2013), however the details of exactly how, why and when they are using specific social applications or reasons for migrating to others applications (or "apps") such as Instagram and Snapchat remain largely unexplored. Pew further reports (Lenhart, 2012) that texting is on the rise for teens and is the "dominant daily mode of communication" (even over phone calls) with 75% of all teens texting as of 2011. While these reports suggest broad hints about some app usage, we wanted to investigate the more nuanced practices of smartphone and app use in the study reported here. We wanted to understand exactly which applications were being used, for how long, and at what times of day in order to complete a much more detailed picture of what teens were doing on their phones than survey-based methods could provide. Additionally, we wanted to pair real-world quantitative data about application use with qualitative methods to gain a deeper understanding of why particular applications were chosen for different types of communication or to different audiences. Finally, we wanted to understand teen practices around sharing, privacy, and managing connections and groups (e.g., unfriending or unfollowing).

Large-scale survey research has shown that teens are among the most rapid adopters of new technologies (Zickuhr, 2010), and teens, as early adopters, can be a strong predictor of more general adoption trends. However, explicitly studying the reasons for new application adoption by teens is underexplored, with most research on teen's use of social networks occurring before the widespread adoption of smartphones, for example (boyd, 2010). To our knowledge, no detailed quantitative analysis of app use for teens, using real data logged from their smartphones, has been conducted. As such, many questions are unanswered about how teens are using these devices in their daily lives, from time and duration of application usage to deeper qualitative understandings of why teens choose specific apps for communication in various contexts.

For example, understanding teens' specific practices for choosing what content to share, when to share it, with whom, and through which app, allows us to explore the types of communication that are currently occurring as well as communications challenges that teens face. This understanding of the desired reach and persistence of communications provides implications for designing new tools that seek to respect teens' preferred boundaries, nuanced sharing practices, and desires around impermanence of content and identity management.

To investigate these practices, we conducted a 14-day mixed methods study with 14 teenagers of diverse backgrounds from 11 cities in the greater San Francisco Bay Area. We combined interviews and twice-daily voicemail diaries about their use of applications with an automated logger that ran on both iOS and Android and kept track of each app that they used, as well as when and for how long it was used. In this paper, we provide quantitative detail on the types of apps that teens used throughout their day as well as detailed qualitative analysis on why our participants chose apps for various types of activities and needs. We report more specifically on communication tasks since this was a significant practice covering two-thirds of both time spent in apps and number of app launches. We will show how the desired persistence and social reach of a particular communication lead to the use of a wide variety of different mobile communications apps.

## Related Work

We began our work by investigating existing studies into mobile application use, teen communication practices, and studies on teens and privacy. These existing studies helped to shape the questions we asked and allowed us to identify differences over time, with specific behaviors evolving rapidly even from a few years ago, as well as differences between teen and adult use of mobile applications.

### Phone Application Use

Several studies have quantitatively investigated mobile app use in the broader population but did not focus on the teen demographic or understanding why certain app usage behaviors emerged. Böhmer et al. (2011) studied temporal patterns of app use among over 4,100 Android phone users and explored the categories of apps that are used at certain times of day, typical app usage durations, and sequences of apps that are commonly used together. The authors highlight interesting temporal differences in app usage as well as the prevalence of communications related apps throughout the day. Overall that study provided a solid understanding of the frequencies of use of particular categories of apps in the broader population and has influenced much of our quantitative analysis of the app usage logs from our teen participants. We will explicitly discuss similarities and differences between our study and the Böhmer et al. (2011) study in our findings below.

Intel Labs explored the idea of *plastic technologies* (Rattenbury, 2008) that fill gaps in between scheduled activities in daily life, such as many mobile phone applications are doing today. We were similarly interested in the extent to which teens were using applications to fill downtime in their lives and in understanding the specific durations of interactions with their phones.

Other studies have explored app usage among particular older demographics, specifically college students. For example, Rahmati et al. (2012) conducted a year-long study of 34 college students using iPhones to investigate how users in different socio-economic groups adopt and appropriate new smartphones. Similar to our approach, the authors use on-device logging software to collect a diverse range of iPhone usage events including app usage, app installation and uninstallation and web usage. Using data collected from Feb 2010 – Feb 2011, the authors highlight college students purchased an average of 14 apps and spent a median of $25 dollars over the course of the study. Yang, Brown, and Braun (2013) also studied college students in 6 geographies to see which media were most useful at different stages of relationship development and how this was influenced by context or gender. This study used a focus-group interview technique and, similar to our approach, a thematic analysis of transcript data to categorize qualitative results. Their results suggest that relationships evolved from new acquaintances using Facebook posts, progressing to IM, followed by phone calls, and finally in person meetings. Their findings suggest, in part, that by virtue of its public nature "Facebook hindered intimate conversation and thus interaction on the platform was casual and superficial"; it was viewed more as a "bulletin board", not a communication channel. Most recently Lee et al. (2014) focused on the negative aspects of smartphone use among college students. Using a mixed-method approach, which combined logging actual app usage with surveys and interviews, the authors investigated if and how app usage relates to smartphone overuse and addiction.

A recent Nielsen report on mobile device use (March, 2014) found *monthly* usage of mobile apps for adults at ~30hours per month (men were 29hrs 32 mins, women 30 hrs 58 mins) representing an increase over 2013 usage stats of approximate +7 hours/month. They additionally reported *weekly* time spent using any mobile app/mobile web for different age demographics but this did not report any data for for 12-17 year olds (data for 18-24yr olds was reported at 7hrs:10 mins per week or ~41 hours per month). (Note that the Nielsen Electronic Mobile Measurement method does not include "making/receiving phone calls, sending SMS/MMS messages, etc.", though perhaps similar to our logging tools it does use "passive metering.")

While studies to date have shed light on general app usage (typically through surveys of mobile phone use), and app usage among college students, little is known about the types of apps now used by teens and more importantly how and why they choose certain apps in particular situations. The goal of our work is to bridge that gap.

*Teen Communication Practices*

Before the development of the modern iOS and Android smartphone, many researchers studied how teens were using more basic phones with phone calling and texting capabilities. Grinter and Eldridge (2003) studied teen use of text messaging in the UK. They found that teens were communicating with "surprisingly few friends" via their mobile phones and that text messages often led to other forms of communication. Also notable, their work reports that teens were often not communicating with multiple people at the same time. We wondered if these practices from 2003 were still consistent a decade later and sought to answer questions around group communication and the number of friends that teens communicated with given the subsequent rise of Social Network Services and more capable smartphones since that time.

Around the same time, Ling studied the use of mobile phones by teens in Norway (2001). Specifically, he explored the use of the mobile phone to micro-coordinate between teens and with parents. Later, but still before the iPhone and Android devices reached the market, Ling (2004, 2005) conducted a survey to explore mobile communication patterns of teens. In particular, Ling differentiated *micro-coordination* typical of coordinating families from young adults hyper-coordination when such "instrumental planning and management are amplified by continuous mediated access and interaction" (2004). While providing some quantitative statistics similar to later Pew studies (e.g., Madden et al., 2013), this work was limited by the survey method and as such was unable to explore communication choices in depth. This study was also conducted when voice calls or SMS were the only options for communication on the phone. We were interested in exploring how practices have changed as well as how current applications are used to perform some of the same coordination tasks that Ling studied in the pre-smartphone era.

boyd (2010, 2011, 2014) studied teen practices on Social Network Sites from 2004–2007. These studies occurred before teens adopted iOS and Android smartphones and her work focused mainly on practices with websites or minimal "dumb phone" tools. She found teens to be "hanging out" online much in the way that they did historically in malls and on the street. Teens gossiped, discussed popular culture, and generally performed the same activities that they would in person. boyd found that "instant messaging, mobile phones, and social network sites are used interchangeably for a variety of friendship-driven practices." We were interested in investigating the extent that this is still true given the proliferation of smartphones and new communications apps, since boyd's studies were conducted before the widespread adoption of the smartphone. Furthermore, since these studies, new categories of mobile apps have emerged including genres where messages disappear (e.g. Snapchat), group video and voice applications (e.g. Viber, Google Hangouts, Facetime, etc.) and other apps that have moved away from the profile-based articulated networks that boyd studied in depth with teens.

Other studies (Ito, 2001; Chen and Katz, 2008; Weilenmann, 2003) have investigated the ways that teens communicated over the phone, before smartphones and mobile apps were developed. Weilenmann (2003) explored how an 18-year-old teen talked about location with friends and family on her mobile phone. Particular findings explored the use of location sharing for coordinating to meet up in person. Ito (2001) explored how the mobile phone was being used to communicate with friends outside the strict control of parents at home, while Chen and Katz (2008) explored how mobile phones allowed college students to be better connected with their parents and family obligations.

Axelsson (2010) reported on a 2007 survey of 88 Swedish young adults (18-24 year olds) use of mobile phones and found that young adults seem to be "in perpetual contact with family, friends, and colleagues, via traditional voice communication but also via text messages". Again, this was a national survey conducted in 2007 prior to many of the newer IM, photo, and video sharing apps now commonplace on smart phones.

More recently, Hall and Baym (2012) applied dialectical theory to better understand and quantify how "the greater use of cell phones to call and text close friends leads to higher expectations that friends will use mobiles for relational maintenance", building on prior work from Baym (2010). While this connectedness may increase some dimensions of relational satisfaction, it also simultaneously may create a sense of overdependence or even entrapment. In fact, their survey findings support that the use of mobile phone calls and text messages increases a sense of obligation to be responsive while simultaneously creating this interdependence and sense of entrapment. Again, this study relied on self-reports via a survey, while our study directly collects quantitative and qualitative data from participants to augment the understanding obtained from these previous studies. We have not found a study that comprehensively analyzes the current practices of how teens are using their smartphones.

## Research questions

We identified a set of research questions that were unanswered by the existing literature, focusing on the broader goal of learning exactly how today's teens were using smartphones in their daily lives. We were particularly interested in knowing what percentage of their time was spent in different categories of apps, by hour of the day, and why they chose particular communications apps to use for specific messages. More broadly, we want to understand the roles of different types of messaging and social networking apps in their lives, why they chose particular apps and did not use (or had stopped using) others. Additionally, we wanted to both confirm prior reports (e.g., Madden et al., 2013; Wortham, 2013) that today's teens have largely replaced Facebook with other apps and to develop an understanding of their practices around unfriending, unfollowing, or cleaning up their contacts or people-based lists. From this, we were not only able to articulate key current practices, but also developed design implications and guidelines for new mobile communications applications.

## Methods

We conducted a mixed-method study with 14 participants of quite diverse backgrounds from the greater San Francisco Bay area to explore how teens aged 13 to 18 used their smartphones in daily life. All participants were existing smartphone users who had used their device for over six months.

The study consisted of an initial in-person interview, twice-daily voice diaries for 14 days describing their smartphone use, and a final in-person interview at the end of the study. Between the initial and final interviews (which averaged about 18 days depending on participant availability), participants also installed and ran a logger on their smartphone that kept track of their application use.

Teens were recruited from a large existing database of potential participants as well as through a professional recruiting firm to increase diversity. In general, we sought the broadest sample possible from the greater San Francisco Bay Area. While the Bay Area tends to be wealthier and more educated than the rest of America, we explicitly recruited for participants from lower income areas. We recruited 5 males and 9 females, half whom used an iPhone and half whom used an Android smartphone. Half of the participants were between 13–15 years old, while half were 16–18 years old. Participants were recruited from 11 different cities around the greater San Francisco Bay Area to represent various ethnic and socio-economic backgrounds. Similar to the U.S. population, 2/3 of participants were selected such that neither parent had a university degree. All participants spoke English fluently although several spoke other languages at home. Parental consent was obtained for all participants under 18. Participants were compensated for their time. While we cannot claim to be representative of the entire U.S. teen population with just 14 participants, the diversity in participants helps us ensure that themes across quite different participants are more likely to be true more broadly in the general population. All qualitative data presented below represents these cross-participant themes.

In the initial interview, we asked participants to list the people that they communicate with and to group these people in categories that the participant defined (similar to Bentley et al. (2010)); this gave us contextual data we could later use to interpret our qualitative voice diary data. Participants could use any groupings or group labels they wanted and contacts could be in multiple groups. We used physical groupings of notecards on a tabletop so participants could easily create new cards, move them, and group them. We then asked participants about their perceptions of which apps they used and how often they used them. We then went through each app and asked who the participant thought their intended audience was for content they posted or sent. We asked who they thought could see this content and how this influenced their choices. We additionally asked about apps that they used a year ago but no longer used now (and why), and we took an inventory of all the apps they currently had loaded on their smartphone.

Finally, we installed a custom software logging tool on their smartphone and showed samples of the data it would collect. Our loggers, built for both iOS and Android, collected time stamped application use data: the name of the app that was currently active, the time it was launched, and how long that app stayed active in the foreground. We did not collect any keystroke data, data about the content shared, or data about who the participants were communicating with. Because the loggers keep track of application use while the display was active, we will be reporting the time that teens spent *interacting* with their phone. For example, if the display turned off while background audio was playing, we would only count the time that the participant spent actively using the music application with the screen lit in the analysis below. This application logger ran as a passive background process on the phone and did not impact app use or other phone use patterns (though it did consume slightly more battery power on iPhone devices).

For the next 14 days, participants were prompted through a phone-based calendar notification to leave us a voicemail message twice per day summarizing how they used their smartphone in the preceding few hours, whom they communicated with, by what means/app, and were asked to detail stories related to their app or device use. Calendar reminders were set to automatically pop up a call-in reminder and enable a single button-press to reach our voice mailbox. Times were set to avoid in-school hours (i.e., after 3:30pm) and the last reminder was set for late evening. During this same 14-day period our application logger (described above) ran in the background on their devices and pushed their app use to our servers daily.

At the end of the logging period, participants attended a final in-person interview. Prior to the interview, we reviewed their application usage log data and their diary entries to list topics or data we wanted to follow up on or get more details for. We asked about any activity from their logged app usage data that was not reported in their voice diary entries. We also asked about "unfriending" or "unfollowing" practices, and if they thought the study has altered their device use. Finally, we asked about their account settings and whether or not their smartphone was password protected to better understand privacy settings or concerns they might have. All in-person interviews were audio recorded and transcribed and all voicemail dairy logs were transcribed.

*Limitations*

While we gathered a large amount of qualitative and quantitative data about teens' use of mobile phones, there are some limitations to the methods that we employed that we would like to highlight. While we explicitly recruited teens from different ethnic and socio-economic backgrounds (such as choosing 2/3 of participants whose parents did not attend college) and different cities, our study was only conducted in one region, the San Francisco Bay Area, which is typically more affluent and educated than much of America and may have cultural differences from other parts of the US or other countries.

Given the two-week nature of our study, we are only seeing a slice of usage (specifically in the Fall of 2013). Our study period captured the start of the school year, while other periods of time might result in different findings. Our app logger missed a day or two of data for several iOS participants who accidently terminated the logger, however, on average we collected 18 days of data per participant based on the time between the initial and final interviews. As mentioned earlier, we are measuring *active* interaction time with the smartphone, thus we did not log entire session durations for long phone calls or music playing sessions where the display turns off and the user has no direct interactions with the smartphone.

*Analysis*

*Quantitative data* was analyzed both per-participant and in aggregate to understand the types of apps that were used, when specific apps were used, and for how long. Direct comparisons to Böhmer *el al.*'s study of smartphone app usage in the general population (2011) is provided where possible since that was the closest prior work. *Qualitative data* from interviews and voicemail diaries was transcribed and put into a 1200-note grounded-theory based affinity analysis (Beyer and Holtzblatt, 1999). Grounded theory-based affinity analysis is an approach commonly used to organize and group large quantities of qualitative data into logical and linked categories. Typically multiple reviewers or judges create a converged set of themes or categories and come to agreement on categorization of the data elements thus providing some inter-judge reliability. In our study, over 1,200 individual quotes were extracted from the transcribed qualitative data, which made up the individual data items in our qualitative analysis (unlike the more rapid Beyer and Holtzblatt (1999) method based on researcher insights). Four researchers then iteratively grouped these data items, and themes were identified across participants as the data was combined. Each of the emergent themes is discussed below and represents data from multiple participants that converged to this theme based on our analysis. Through this combination of quantitative and qualitative data, we were able to get a rich picture from multiple data sources to see what teens do with their smartphones and, perhaps more importantly, why they make decisions to use certain apps in particular situations or with particular social groups.

# Findings

In combining our qualitative findings with the quantitative data, we wanted to understand several key issues. Are there temporal patterns that reveal interesting times of day when some apps are used more? Which apps are most used/popular and which categories of apps are most used? The results reported in this section are based on the usage logs gathered between August 19[th] and September 23[rd], 2013. (Although the diary part of the study lasted exactly 14 days, due to scheduling interviews, the loggers ran for an average of 18 days.) We logged a total of 75,004 mobile phone usage events over the entire study time (avg 5,357 events per participant, sd: 4,085.7[1]), of which 32,524 were app launches (avg 2,323 per participant, sd: 1,964.6). The remaining usage events were events such as locking/unlocking their device and accessing the homescreen of their phone. Our 14 participants used a total of 258 unique mobile apps over the course of the study.

---

[1] Note that the average number of events over the entire study has an expected large standard deviation representing the high variance between subjects and even between days within subjects. This number is mentioned to present an overview of the scale the data set rather than a specific research interest in an aggregate number of events over 14-18 days.

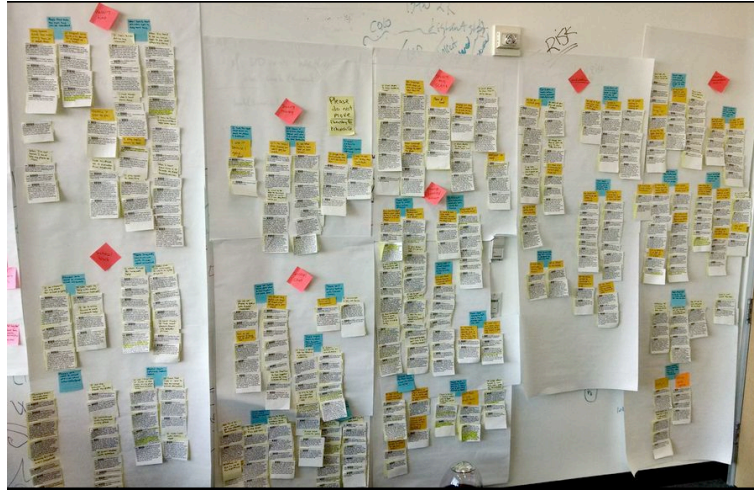

Figure 1: One wall of our grounded theory-based affinity analysis of our qualitative data. All white notes are exact quote from our participants. Yellow and blue notes are emergent themes or categories.

## Overview of apps used

Overall, we found that the total time of active engagement with their phones for the teens in our study is significantly higher than data from the general population of all smartphone users at the time the study was conducted. One Nielsen study (2009) found that the general population spent an average of 39 minutes per day in apps, a second more recent Nielsen report (2014) found use of 1 hour per day, while Böhmer et al. reported 59 minutes per day of total use (2011). In contrast, aggregated over all participants, our teens spent an average of 173.6 minutes (almost 3 hours, sd: 129.7 mins) per day interacting with their smartphone and 130.9 minutes (over 2 hours, sd: 79 mins) per day using mobile apps. The difference between these two numbers includes time spent with the phone screen lit, but not in a particular application – such as looking at the clock, using the homescreen/launcher, or entering an unlock code. (Recall that these teen participants are in school for much of the day and often have many scheduled after-school activities). Despite spending large amounts of overall time using their smartphones each day, individual mobile app usage is often short or "bursty," lasting an average of 63 seconds (sd: 174.6 seconds), slightly shorter than Böhmer et al.'s 71 second average for the general population. Overall we find that 77.7% of app uses were one minute or less in duration, which provides evidence that the short, bursty nature of mobile app use is also present among teens, consistent with findings for the general population, from both (Bohmer, 2011) and Nielsen (2009, 2014).

On average, our participants each used 36.6 unique apps during the study (min:22 apps, max:64, sd:12.9). We logged an average of 123.5 app launches per participant per day (min:37 apps, max:387.3, sd:97.1) and found an average of 15.6 categories of apps per participant (min:12 app categories, max:22, sd:2.7). To get a better sense of the types of apps our teen participants used, we manually classified all 258 unique apps into one of 26 categories. To do so we analyzed both Google Play and the iOS app stores, searching for each app's category by app name. While most app categories are consistent across Google Play and the iOS app store, some changes were required to accommodate our diverse dataset. For example, instead of having multiple micro-categories of *Games* (e.g. arcade, brain & puzzle, etc.) we opted for one high-level games category. Similar to Böhmer et al. (2011) we opted to have a separate *Browsers* category as well as a separate *Systems & Settings* category for handling the default Android and iPhone settings apps. Finally, given that we were particularly interested in understanding how teens used their phones for communication and because of its predominance in our data, we opted to break-out communications-related categories such as *Contacts*, *SMS/Texting* and *Instant Messaging (IM)*. Note that we also reviewed the app categorization schemes used by comScore (2013), Pew (Madden et al., 2013) and Nielsen (2009) but found these to be overly complex, or in the case of Pew and Nielsen to not discriminate sufficiently between the various communications apps such as SMS and IM.

| Category | # App Launches | % App Launches | # Apps | Avg. Dur (sec) | Examples |
|---|---|---|---|---|---|
| SMS/Texting | 6901 | 21.2 | 3 | 72.0 | Built-in SMS apps |
| Social networking | 6302 | 19.4 | 18 | 107.8 | Instagram, Facebook, Tumblr |
| Phone & Audio Comms | 3693 | 11.4 | 9 | 28.2 | Built-in dialer, Google Voice, Voxer |
| Instant Messaging | 2711 | 8.3 | 10 | 162.6 | Snapchat, Kik, TextNow |
| Music & Audio | 1829 | 5.6 | 15 | 95.4 | Pandora, build-in music player |
| System & Settings | 1566 | 4.8 | 10 | 14.7 | Chrome, Firefox |
| Browsers | 1531 | 4.7 | 8 | 97.1 | Built-in settings, preferences |
| Productivity | 1478 | 4.5 | 31 | 33.1 | Calendar, Gtasks, Evernote |
| Photography | 1296 | 4.0 | 21 | 102.4 | Built-in gallery & camera, instacollage |
| Tools & Utilities | 1231 | 3.8 | 37 | 97.3 | Clock, alarm, flashlight, app store |
| Contact | 1133 | 3.5 | 2 | 42.1 | Gmail, Y! Mail |
| Email | 978 | 3.0 | 6 | 31.7 | Default / built-in contact app |
| Media & Video | 627 | 1.9 | 16 | 191.3 | Vine, Youtube, Netflix |
| Games | 556 | 1.7 | 32 | 120.7 | Candy Crush Saga, Fun Run |
| Entertainment | 188 | 0.6 | 1 | 271.9 | iFunny |
| News & Magazines | 138 | 0.4 | 4 | 171.7 | Flipboard, Reddit, Yahoo! News |
| Shopping & Retail | 110 | 0.3 | 8 | 85.5 | Poshmark, ebay, Amazon |
| Travel & Local | 98 | 0.3 | 6 | 134.4 | Google maps, atlas, phone tracker |
| Weather | 57 | 0.2 | 2 | 36.9 | Built-in weather apps |
| Books & Reference | 50 | 0.2 | 7 | 159.2 | iBooks, Urban Dictionary, Wikipedia |
| Finance | 27 | 0.1 | 2 | 33.6 | Wells Fargo, Bank of America |
| Unknown | 6 | 0.0 | 3 | 12.2 | NA |
| Personalization | 6 | 0.0 | 3 | 39.3 | Zedge, Live wallpaper picker |
| Health & Fitness | 5 | 0.0 | 1 | 65.2 | P Tracker |
| Education | 5 | 0.0 | 1 | 106.2 | Canvas |
| Lifestyle | 2 | 0.0 | 2 | 5.0 | Nissan LEAF, Vivino |

Table 1: Number and percentage of app launches, number of distinct apps and associated app durations (average in seconds) per app category. (ordered by number of app launches).

From Table 1, we found *shorter* average usage times for app categories like *Lifestyle, Unknown*[2], *Contact List, Phone & Audio Communication*, and *Email*. We found *longer* average usage durations for apps related to *Games, Travel & Local, Entertainment, Media & Video* and *News & Magazines*. Longer sessions times for video and media has been found in other studies (Nielsen, 2014), for example those looking at mobile phones and tablets. Note that the very short average usage time for the *Lifestyle* category is due to the fact that there were only 2 apps found within this category both used once, very briefly.

Overall we find that the majority of app usage logged during our study relates to *SMS/Texting* (21.2% of launches), *Social networking* (19.4%) and *Phone & Audio Communications* (11.4%). If we consider all related categories, specifically *Instant Messaging* (8.3%), the *Contact List* (3.5%) and *Email* (3.0%) we find **communication** represents 66.8% of all app launches. Note that our findings for IM, SMS, texting are about 30% of total app launches while phone calls represent approximately 11% of app launches, consistent with the Pew report findings (Lenhart, 2013) that teen texting far surpasses phone calls as a means of communication.

Awareness of Total App Use
Teens often found themselves drawn to their phones in any spare second, whether waiting for parents to pick them up, waiting for friends to show up, or just when having unscheduled time. In these instances, they turned to their smartphones for entertainment and something to do.

Games were often played while waiting. P15 was at an amusement park with friends and decided not to go on one of the bigger roller coasters. She waited for them while playing the game Mine Runner. P13 plays games when she walks home from school alone: "*I'm walking home alone so it's kind of when I'm bored. I'd rather play games or text people.*" Movies and video clips were also watched when passing time. P2 said that "*Netflix is fun you know. In case you get bored.*" P3 told us that she likes watching movies or playing Candy Crush when she's sick and stuck in bed.

---

[2] Examples of specific apps in the Unknown category include: DirectShareManager and Multimedia UI Service Layer. Apps were assigned to this category when we could not find enough information about them to assign them to an alternative category.

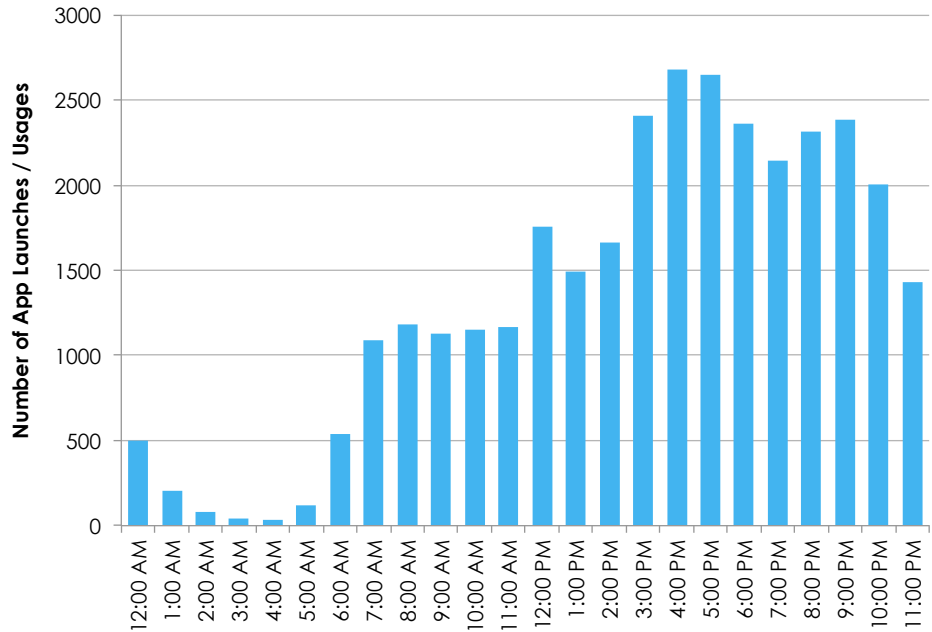

Figure 2: Number of app launches during a day

Smartphones were also used in long car rides with parents to and from school. P10 used her smartphone "*on the way home. Whenever I'm in the car and I'm bored.*" She also talked about listening to music during these long car rides and P12 made a point of putting in headphones in the car, and thus not engaging with her parents. This can be seen as a way of teens trying to be Alone Together (Turkle, 2012) with their parents by connecting to distant friends instead of interacting with parents who were sitting right next to them.

However, we also saw that teens were quite aware of the amount of the time that they spent on their smartphones. To reduce distractions, they often removed apps from their smartphones that they felt were the biggest time wasters. P1 "*deleted almost all of my games so I could stay focused.*" And P11 "*never played [Candy Crush] just because I was tempted to, but my roommate lost her life doing that. She's on it all the time. … They get very addicting and you just keep doing it!*" P1 stopped using Snapchat because "*it got distracting so I just told myself to stop. Like not getting work done, not being productive.*"

Overall, smartphones provided a welcomed distraction for teens when they were bored. From music, to games, to social updates, there were many apps to keep them occupied. However, several participants were quite aware of the addicting power of some apps, and chose to remove (or never install) ones that were seen to be the biggest distractions.

App Usage Over Time
Unlike Böhmer et al. (2011), we do not find consistent linear growth throughout the course of the day when it comes to the total number of app launches, as shown in Figure 2. Instead we find peaks of application usage around noon, between 3–5pm (just after school), which is when total application usage reaches its maximum, and another slight peak around 8–9pm. Usage generally begins between 6–7am. Even though many of our participants' high schools had "devices off" rules, there is use throughout the day including during school hours. In Figure 3 we separate out weekday use from weekend use by time of day. In contrast, we observe an increase in use in the mid-mornings and around lunch time on weekends for teens with less usage in the early mornings (perhaps in part because they are not in school and teens are known to sleep in on weekends). Otherwise, the usage pattern seems roughly consistent between weekends and weekdays. Studies of adults (e.g., Muller et al., 2012) have shown drops in mobile device and tablet use on weekends versus weekdays.

When we examine app categories, we find differences in the relative use between different categories over the course of the day. Table 2 shows a heat map visualization of the launch frequency of each app category by time of day. Given the prevalence of communication usage, we separate communication-related apps (blue at top of Table 2) from non-communication-related apps (highlighted in orange), with darker shades representing higher percentages off app launches in those time slots.

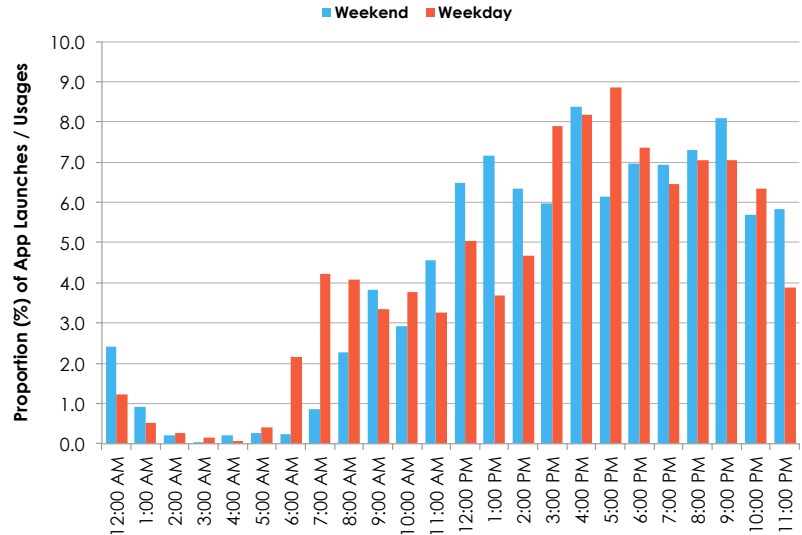

**Figure 3: Proportion of app launches / usages during a day comparing weekends and weekdays**

Overall this heatmap data shows that use is sprinkled throughout the day for many application types with peak times immediately following the end of school (about 3pm–5pm) or just prior to bed time (10-11pm). However, for certain app categories there are stronger temporal patterns. For example, *Weather* related apps are used more in the mornings (between 6–7am). *Media & Video* occurs more at night, in particular between 10-11pm (consistent with findings from Neilsen, 2014). We find that usage of *Email*, *SMS/Texting* and *Instant Messaging* was more consistent throughout the day with peaks around 4–5pm and again between 8–9pm. *Phone & Audio Communication* is used most often between 3–4pm. Given that many of the teens used phone calls as a means of coordinating with their parents and friends, these peaks tend to correspond to when school finishes and they either call friends to coordinate or their parents come to pick them up either at school or at a nearby train station.

| | 12AM | 1AM | 2AM | 3AM | 4AM | 5AM | 6AM | 7AM | 8AM | 9AM | 10AM | 11AM | 12PM | 1PM | 2PM | 3PM | 4PM | 5PM | 6PM | 7PM | 8PM | 9PM | 10PM | 11PM |
|---|---|---|---|---|---|---|---|---|---|---|---|---|---|---|---|---|---|---|---|---|---|---|---|---|
| Contact | 0.4 | 0.0 | 0.0 | 0.0 | 0.0 | 0.0 | 0.3 | 1.1 | 4.2 | 2.7 | 2.3 | 1.3 | 4.3 | 3.2 | 5.0 | 15.7 | 16.2 | 9.0 | 6.7 | 7.2 | 6.1 | 7.9 | 4.1 | 2.1 |
| Email | 1.8 | 1.2 | 0.5 | 0.1 | 0.2 | 0.2 | 3.4 | 5.8 | 4.2 | 3.1 | 5.2 | 2.9 | 4.6 | 4.3 | 5.5 | 7.4 | 10.1 | 7.5 | 4.8 | 5.3 | 9.1 | 6.4 | 3.9 | 2.5 |
| Instant Messaging | 0.6 | 0.1 | 0.1 | 0.0 | 0.0 | 0.0 | 0.5 | 1.4 | 2.5 | 2.1 | 3.6 | 4.3 | 4.7 | 5.2 | 7.6 | 7.3 | 5.8 | 10.4 | 8.9 | 4.9 | 7.6 | 10.5 | 8.3 | 3.6 |
| Phone & Audio Comms | 2.8 | 1.5 | 1.2 | 0.1 | 0.2 | 0.3 | 1.2 | 3.3 | 2.2 | 5.2 | 3.5 | 2.9 | 6.7 | 3.9 | 2.1 | 2.5 | 8.3 | 7.5 | 7.5 | 6.9 | 9.0 | 8.6 | 6.6 | 5.9 |
| SMS/Texting | 1.3 | 0.3 | 0.2 | 0.1 | 0.1 | 0.3 | 0.6 | 3.1 | 3.3 | 2.8 | 2.7 | 3.8 | 5.3 | 4.6 | 6.1 | 7.1 | 7.2 | 7.4 | 8.9 | 7.8 | 8.1 | 8.4 | 6.5 | 4.0 |
| Social networking | 2.8 | 1.1 | 0.2 | 0.2 | 0.1 | 0.3 | 2.1 | 3.9 | 2.7 | 4.0 | 3.6 | 3.4 | 4.6 | 4.0 | 3.7 | 4.8 | 7.8 | 8.3 | 6.5 | 7.0 | 8.2 | 6.5 | 7.9 | 6.4 |
| Books & Reference | 0.0 | 0.0 | 0.0 | 0.0 | 0.0 | 0.0 | 4.0 | 2.0 | 0.0 | 0.0 | 0.0 | 0.0 | 6.0 | 0.0 | 0.0 | 0.0 | 26.0 | 2.0 | 10.0 | 2.0 | 44.0 | 2.0 | 2.0 | 0.0 |
| Browsers | 1.6 | 0.1 | 0.6 | 0.7 | 0.3 | 0.3 | 1.4 | 3.2 | 5.1 | 6.1 | 5.7 | 5.4 | 6.1 | 6.4 | 6.4 | 4.8 | 6.9 | 8.0 | 3.7 | 6.3 | 5.4 | 6.2 | 5.7 | 3.6 |
| Education | 0.0 | 0.0 | 0.0 | 0.0 | 0.0 | 0.0 | 60.0 | 0.0 | 0.0 | 0.0 | 0.0 | 0.0 | 0.0 | 0.0 | 0.0 | 0.0 | 0.0 | 0.0 | 0.0 | 0.0 | 40.0 | 0.0 | 0.0 | 0.0 |
| Entertainment | 0.5 | 0.0 | 0.0 | 0.0 | 0.0 | 0.0 | 1.6 | 20.7 | 10.1 | 5.3 | 2.7 | 2.1 | 3.7 | 5.9 | 1.6 | 3.7 | 5.9 | 8.0 | 8.0 | 5.9 | 9.0 | 2.1 | 3.2 | 0.0 |
| Finance | 0.0 | 0.0 | 0.0 | 0.0 | 0.0 | 0.0 | 0.0 | 0.0 | 7.4 | 0.0 | 7.4 | 7.4 | 7.4 | 18.5 | 7.4 | 7.4 | 18.5 | 11.1 | 0.0 | 3.7 | 3.7 | 0.0 | 0.0 | 0.0 |
| Games | 0.7 | 0.2 | 0.0 | 0.2 | 0.2 | 0.0 | 0.9 | 2.2 | 5.4 | 4.7 | 6.8 | 5.0 | 5.4 | 7.6 | 4.3 | 3.8 | 2.2 | 9.9 | 7.6 | 9.2 | 7.4 | 5.9 | 8.1 | 2.5 |
| Health & Fitness | 0.0 | 0.0 | 0.0 | 0.0 | 0.0 | 0.0 | 0.0 | 0.0 | 0.0 | 0.0 | 0.0 | 0.0 | 0.0 | 0.0 | 20.0 | 20.0 | 20.0 | 20.0 | 0.0 | 0.0 | 0.0 | 0.0 | 0.0 | 20.0 |
| Lifestyle | 0.0 | 0.0 | 0.0 | 0.0 | 0.0 | 0.0 | 0.0 | 0.0 | 0.0 | 0.0 | 50.0 | 0.0 | 0.0 | 0.0 | 0.0 | 0.0 | 0.0 | 0.0 | 0.0 | 0.0 | 0.0 | 0.0 | 0.0 | 50.0 |
| Media & Video | 4.5 | 1.0 | 0.0 | 0.8 | 0.0 | 0.0 | 1.0 | 4.1 | 1.1 | 2.2 | 1.9 | 3.0 | 4.9 | 3.3 | 1.6 | 7.3 | 7.5 | 7.2 | 8.0 | 9.1 | 3.0 | 6.1 | 11.2 | 11.2 |
| Music & Audio | 0.4 | 0.2 | 0.1 | 0.0 | 0.0 | 0.1 | 1.1 | 3.9 | 6.9 | 4.1 | 4.6 | 7.0 | 9.7 | 7.5 | 7.3 | 8.6 | 5.6 | 6.1 | 5.7 | 4.9 | 4.7 | 4.5 | 1.5 | |
| News & Magazines | 2.9 | 0.0 | 2.2 | 0.0 | 1.4 | 0.0 | 5.8 | 6.5 | 2.2 | 1.4 | 1.4 | 3.6 | 5.8 | 3.6 | 6.5 | 3.6 | 9.4 | 13.8 | 4.3 | 2.2 | 6.5 | 2.2 | 6.5 | 8.0 |
| Personalization | 0.0 | 0.0 | 0.0 | 0.0 | 0.0 | 0.0 | 0.0 | 0.0 | 0.0 | 0.0 | 66.7 | 0.0 | 0.0 | 0.0 | 0.0 | 0.0 | 0.0 | 0.0 | 0.0 | 33.3 | 0.0 | 0.0 | 0.0 | 0.0 |
| Photography | 2.8 | 1.5 | 1.2 | 0.1 | 0.2 | 0.3 | 1.2 | 3.3 | 2.2 | 5.2 | 3.5 | 2.9 | 6.7 | 3.9 | 2.1 | 2.5 | 8.3 | 7.5 | 7.5 | 6.9 | 9.0 | 8.6 | 6.6 | 5.9 |
| Productivity | 1.5 | 0.7 | 0.1 | 0.0 | 0.5 | 0.1 | 2.4 | 2.6 | 2.0 | 1.8 | 3.4 | 2.1 | 3.2 | 2.7 | 4.9 | 14.7 | 9.6 | 6.7 | 4.4 | 5.8 | 11.0 | 11.8 | 4.5 | 3.2 |
| Shopping & Retail | 4.5 | 7.3 | 8.2 | 0.0 | 1.8 | 2.7 | 1.8 | 0.0 | 3.6 | 3.6 | 1.8 | 5.5 | 0.0 | 0.0 | 3.6 | 3.6 | 2.7 | 4.5 | 4.5 | 1.8 | 1.8 | 13.6 | 11.8 | 10.9 |
| System & Settings | 1.7 | 0.6 | 0.0 | 0.0 | 0.1 | 0.1 | 1.6 | 3.2 | 6.3 | 4.7 | 5.6 | 4.2 | 6.5 | 6.6 | 5.7 | 5.2 | 5.7 | 9.1 | 7.5 | 5.2 | 4.9 | 5.9 | 4.5 | 5.2 |
| Tools & Utilities | 2.0 | 1.6 | 0.1 | 0.1 | 0.2 | 5.0 | 11.9 | 8.4 | 3.2 | 1.9 | 3.6 | 2.1 | 3.1 | 1.5 | 2.4 | 3.7 | 3.6 | 8.4 | 9.0 | 4.9 | 5.2 | 3.1 | 6.9 | 8.2 |
| Travel & Local | 3.1 | 0.0 | 0.0 | 0.0 | 0.0 | 0.0 | 1.0 | 8.2 | 5.1 | 10.2 | 0.0 | 1.0 | 2.0 | 3.1 | 4.1 | 15.3 | 4.1 | 9.2 | 13.3 | 11.2 | 6.1 | 1.0 | 0.0 | 2.0 |
| Unknown | 16.7 | 0.0 | 0.0 | 0.0 | 0.0 | 16.7 | 0.0 | 0.0 | 0.0 | 0.0 | 16.7 | 0.0 | 0.0 | 0.0 | 0.0 | 0.0 | 0.0 | 0.0 | 0.0 | 33.3 | 0.0 | 0.0 | 0.0 | 16.7 |
| Weather | 0.0 | 0.0 | 0.0 | 0.0 | 0.0 | 0.0 | 26.3 | 22.8 | 0.0 | 5.3 | 1.8 | 0.0 | 3.5 | 0.0 | 1.8 | 0.0 | 1.8 | 0.0 | 1.8 | 12.3 | 1.8 | 10.5 | 5.3 | 5.3 |

Table 2: Percentage of app launches per app category by hour of day. Communications-related categories are in blue. Darker colors represent a higher percentage of app launches.

### Rapid App Migration and Reformulation of Friend Groups

While participants used an average of 36.6 different applications over the course of our study, these did not stay consistent over the long term. The apps that teens used quickly came into and out of favor, which led to issues in coordination. Instead of carefully adding to or removing/unfriending people on existing social networks, our data indicated that teens quickly migrated to new apps and re-created groups of desired friends there. The old groups simply got left behind to go inactive without any specific list management. P14's friends were already starting to move off of Snapchat onto other apps. P13 discussed getting on Facebook and then not using email and now moving to Twitter instead of Facebook. Facebook "non-use" or abandonment was previously studied by Baumer et al. (2013) yet they left the study of this behavior in teens as future work. Our participants said they rarely or never "cleaned up" contacts lists or unfollowed or unfriended people (with the possible exception of some celebrity sites). This suggests that some of social stigma reported in the popular press about "unfriending" may be leading to evolving practices to avoid these types of awkward interactions. What we did notice was that these app migrations were often for small groups of close friends. Teens commented that when someone who was just an acquaintance at school, but not a close friend, posted a "like" on their content, it was "weird" - implying that person was not part of the smaller group that should interact in that way. This rapid migration to new apps (or new accounts) also means that parents cannot keep up on which app their teens are using and thus preserves a sense of freedom from monitoring for the teen.

This rapid migration can lead to a problem of knowing how to contact particular people. P10 said that "*most people who have Kik just use that to IM ... The people who don't have it usually just text me.*" Different "friend groups" were often using different communication apps (e.g., school friends versus neighborhood friends or friends from a sports team). Texting seemed to be the common default that would work with almost everyone. Texts were delivered quickly and were seen as the most reliable to broadly cover friends with different phone models and OS versions regardless of which apps were installed.

Games were another category where specific apps came in and out of favor. During the course of the study games such as Candy Crush and Temple Run were becoming quite popular and participants discussed installing these while giving up on older games that they had tried.

### Communications Applications

If we focus on communication-related apps we find that teens used an average of **9.4 distinct communications related apps** (min:6 apps, max:16, sd:2.7) during the study period. Figure 4 highlights the prevalence of communications related app usage by showing the proportion of app usage for each specific type of communications application for each participant in our study. On average, almost two-thirds of their time using the smartphone was spent in communications applications. However, for some teens in our study (e.g. P1 and P12) this was as much as 85% of their smartphone app usage.

Table 3 shows the top 10 individuals apps used by our teens. These top 10 apps account for ~65% of all logged app launches. The top two apps, SMS and Phone Calls account for over 30% of all app usage, thus highlighting that our teen demographic still heavily uses the more traditional phone communication tools, while also adopting many new ones.

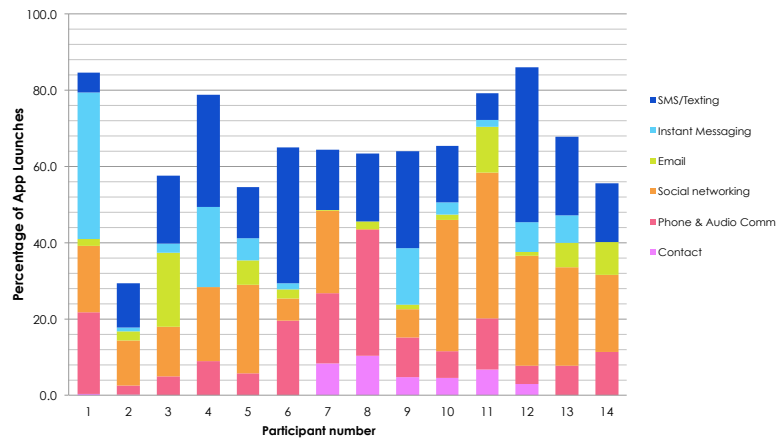

Figure 4: Proportion of communications related app launches/usages per participant.

| App name | # Launches | % Launches |
|---|---|---|
| SMS | 6901 | 21.2 |
| Phone | 3324 | 10.2 |
| Instagram | 3150 | 9.7 |
| Snapchat | 1649 | 5.1 |
| Twitter | 1222 | 3.8 |
| Facebook | 1211 | 3.7 |
| Contacts | 1133 | 3.5 |
| Email | 978 | 3.0 |
| Pandora | 786 | 2.4 |
| Calendar | 774 | 2.4 |

Table 3: Top 10 individual app launches across participants.

We also explored how duration of app usage related to each type of communication application. Figure 5 shows the distribution of application session length as an overall proportion of its total usage[3]. We observe that the vast majority of communications app use is in less than 1-minute sessions, with the exception of some social network apps. Almost half of the phone & audio communications use was less than 5 seconds in length. And overall, app use rarely exceeded 5-minute sessions within a particular application. This is consistent with prior research (Bohmer et al., 2011, Nielsen, 2009) that indicates short, bursty "information snacking" behavior is typical on smartphones. While we confirm this bursty pattern of app use among teens, we observed a significantly higher frequency of app use for a longer period of time throughout the day than was found in the general population in (Bohmer et al., 2011). The duration patterns in Figure 5 were similar for weekends and weekdays.

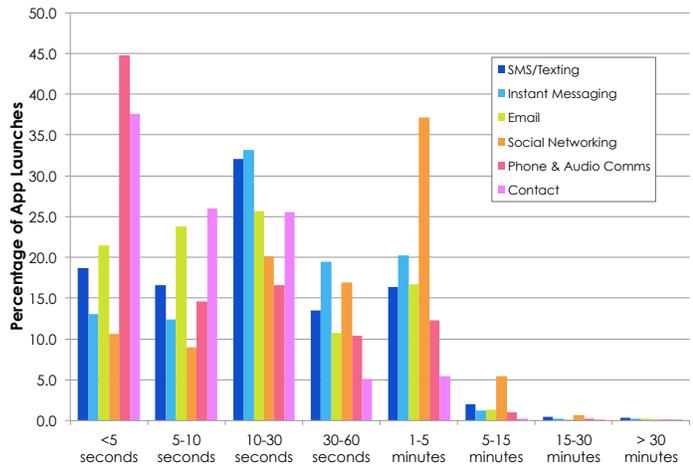

Figure 5: Distribution of app usage based on duration of app sessions for communications related app launches.

Managing Identity and Audience

We observed that smartphones are a critical device where our participants created and shared content about their lives. The teens in our study adopted numerous communications apps that were either synchronous or asynchronous, created a lasting message or disappeared after a few seconds, and supported a range of media including live video/audio, photos, emoticons, and text messages. These applications allow teens to reach out to different groups of people and present themselves in ways that they consider appropriate depending on the chosen audience. We observed that teens had a finely nuanced view of the differences between communications apps and made deliberate choices about the types of content posted on different apps while actively maintaining specific identities and audiences on each service.

---

[3] Note that our logger did not track applications when the display was off, so we are not counting duration for "phone & audio comms" events where no touch events occurred and the display stayed off.

Teens worked quite hard to craft their online personas for sites that had persistent or lasting content. Services that maintained a history of posts, such as Facebook, were seen as potential problems in the future if one was not careful about the content that is posted or tagged with their identity. This is consistent with findings from Gardner and David (2013) which they term the "packaged self: "[w]e gathered considerable evidence that youth take care to present a socially desirable, polished self online" (p.63) and these "glammed up versions of themselves… don't include every detail about their lives" (p.64). Exemplary from our study, P2 told us, "*I don't use Facebook, because I think it can lead to too many problems in your life. You post one thing on there, it's never going to go away. … It can be used against you, and that's one thing I never want to happen to me. I don't want to be looking for a job, and have an employer review my Facebook account and say, hmm, well they posted this ten years ago, that's not good, we don't want to hire them.*" P11 also talked about job issues related to content posting and chose to have a private Twitter account as a consequence. Gardiner and Davis [2013, p. 63] describe how interwoven teens online lives are with their offline lives – "today's young people seldom make a distinction between their online and offline selves". While teens represent that they are "not two different people", we have found considerable evidence that teens present a very "polished self" to wider audiences online.

We saw teens making a clear distinction between places where they communicated with family and zones for different types of friends. P6 told us "*I'm a teenager. I don't tell my mom everything I do. I need to live and learn about stuff so I don't tell my mom everything.*" Facebook was often seen as a zone for family members, especially members of the older generation such as parents, aunts, and uncles. Consistent with other studies (Madden et al., 2013; Wortham, 2013) we also found teens mention that "*Facebook is for like older people… I don't really use it except for school stuff.*" P7 called Facebook "*more of a family thing*" than a place where he communicated with friends. P4 uses Facebook with her aunts and uncles. Teens often had the need to be careful about any Facebook activity with friends, since their parents might see it. P6 told us "*You just think it's funny or want to like it but you know you can't because my mom will see.*" This meant that in practice, communication with friends migrated to other apps such as Instagram or Snapchat, which most participants' parents did not yet use. As we discuss later, Facebook has also been adopted by schools and clubs and thus was also perceived to be for "*more official stuff.*" Taken together, the combination of persistent content and more "*official*" uses, teens often perceived both Facebook and Twitter to be more for public postings to a wide audience that could include people that were acquaintances but were not part of their "inner circle" of closest friends (similar to Yang et al.'s (2013) "bulletin board" characterization with older college students).

Participants discussed putting a great deal of effort into creating photos that they wanted to share on sites that maintain a history, making sure that everyone looked their best. Several participants talked about using special photo collage software to combine multiple images together before posting. P12 posted a collage on Instagram of her softball team after practice: "*We all got really dirty at that practice, we were sliding and everything, so we all took pictures and then posted and made a collage.*" She did this collaboratively with her friends and they told her, "*No, don't post that one. That one is horrible,*" until they found a set that everyone agreed was suitable to post. Participants felt a strong pressure to post content that will be "*liked*" by others when sharing to larger networks. P11 told us that "*It's like stressful, and you feel awkward if no one likes your status or it gets ignored.*" In some cases, content was later removed if it did not immediately attract enough likes or comments.

P12 discussed liking the ability to be artistic on Instagram. She posted a photo from her grandmother's house: "*My grandma has a really cool bathroom. Like she's got a ton of mirrors … you know those mirrors when it looks like a tunnel of you? That's what it looked like and I just felt like the art scene and just took a picture and then like … I posted a picture on Instagram and saying a quote from [a Justin Timberlake] song. … It's hard to explain. My artsy moment.*" Her friends appreciated this "*artsy moment*" and commented on how they also loved that song. One friend hashtagged the photo "#artsymofo," which P12 enjoyed.

This careful curation was contrasted by the type of content that was shared through apps that do not keep a history such as Snapchat, where a photo is directly shared with one or a small number of contacts and the message is automatically deleted after 10 seconds (upon the recipient opening it). P11 told us that "*Snapchat is so temporary that I don't care what the picture is of. … My closest friends, we would send each other really ugly pictures of ourselves, like making a funny face or doing something stupid.*" She also told us that she avoids posting content to Facebook "*because everybody sees it, literally everybody!*" again speaking to this "bulletin board"-like quality of Facebook.

An interesting phenomenon that we observed was "*mass-snapchatting*" where teens would send a photo out to a large portion of their contacts on Snapchat. P11 told us "*I'm not directing a specific message to somebody. … Yesterday I was packing and I knew it was a mess and I just took a picture of it and I was like "Packing for LA" and I just sent it to everybody*." P12 also reported sending a Snapchat to "*all of my contacts.*" This is an interesting new practice where teens do not necessarily want specific messages or photos to be part of a lasting social network profile, but still want to share their status to a wider group of friends.

We also observed participants sending a Snapchat with "*random*" photos in order to have a text conversation that disappears. P7 told us that 75% of his Snapchats are text-based instead of focused on the image. P14 talked about taking pictures of "*mundane things like my foot*" when replying to a Snapchat because the app requires a photo with each message. Thus Snapchat was essentially co-opted to act like an auto-delete IM application (with messages that do not persist) rather than being about sharing the photos themselves.

Other forms of communication were only used for more formal messages. Email fell into this category for many participants. P9 told us that "*I only use email when I need to talk to … like when I'm applying for a job or something that has to do with my school. I don't email my friends.*" P6 said that email "*was like what I use for talking to my coaches and family and stuff.*" P11 called email "*all formal*" while P1 called it "*strictly business.*"

| Application | Total Launches | Average Minutes Per Day |
|---|---|---|
| Snapchat | 1649 | 8.26 min |
| SMS | 6901 | 24.94 min |
| Facebook | 1211 | 11.46 min |
| Instagram | 2900 | 27.19 min |

Table 4: Average time per user per day spent in a variety of communications applications.

Overall, teens had complex reasons for choosing one service over another to share a particular update or photo. As shown in our quantitative results, teens used almost 10 distinct communications applications during the two-week period. Most notably, as seen in Table 4, this new category of ephemeral communication through Snapchat accounted for more total launches than Facebook (1649 vs. 1211 app launches). While Facebook and Instagram, both profile-based services, were still heavily used, this new category of disappearing messaging apps is clearly taking hold. For a new way to communicate, teens are quickly seeing benefits to ephemeral communication tools. The potential audience as well as the duration for which the content would be available to that audience were significant determining factors in choosing an application, as well as the image of themselves that would be portrayed by sharing a piece of content to a given group.

Coordination

Much of teens' lives revolve around coordination or even *hyper-coordination* (Ling, 2005). Not only do they frequently meet up with friends and classmates to socialize, play sports, and work on homework, but they also need to commute to school each day and often need to be picked up or dropped off at activities by their parents.

Past work on SMS behaviors (Grinter and Eldridge, 2003) and emerging mobile instant messaging practices (Church and Oliveira, 2013; Ling 2004; Davidoff et al., 2011; Gardner and Davis, 2013) highlights that one of the key ways that text-based mobile communication is used is to coordinate with other people. Similarly, we found that teens generally preferred text-based forms of coordination when planning to meet up with others. Often, they would use group SMS capabilities to text all of the friends that they were planning on meeting. P13 said, "*If it's concerning a group of people, what we're going to do, we group text it.*" P2 texted a group of teammates about their practice times. Texting was also used to coordinate rides from parents: "*When I'm going home I'll text my mom that I'm on Bart [a regional train in the Bay Area] because she picks me up from the Bart station*" (P15).

This coordination was also visible during sports or club-related activities. P13 talked about coordination after her Color Guard (a group of choreographed flag bearers who accompany a marching band) performances. "*I'll probably text my dad like, 'Hey, halftime is over, I'm going to go get my suit now and then I'll come home in like a little bit.' Then I'll walk over and text my friends, 'Where are you? I don't see you,' that sort of thing.*"

However, we observed that texting is often seen as "too slow" when needing instant coordination or when describing the exact location where one is standing in order to meet up. In these cases, phone calls were often placed. P1 talked about only placing phone calls when he has "*an objective to complete, like, 'Where are you?'*" when trying to find parents to get a ride. P15 also talked about calling if she gets to the Bart station and couldn't find her mom's car. Short phone calls served as a fallback when texting did not successfully complete the coordination task in time.

Some sort of coordination was often needed before synchronous activities began, such as remote gaming, watching the same TV shows, or setting up group Skype calls. Much of this coordination also occurred over text messaging. P11 discussed a friend texting her to watch the same TV show at the same time: "*She would be like, 'You have to see this. Watch it now!'*" Other participants discussed messaging to coordinate group gameplay. P14 discussed sending messages to friends, such as "*Hey guys, get on!*" to get his friends on Xbox together. P7 told us how she would use the Twitter app to see updates from gaming celebrities when they were hosting e-sports tournaments: "*on Twitter they'd say, 'Oh I'm streaming on Twitch right now.'*"

### Gossip/Chat

It is no surprise to discover that teens spent a lot of their time chatting with each other and gossiping about fellow classmates or celebrities (consistent with boyd, 2010). From emerging apps such as ask.fm, to more "traditional" forms of communication such as texting, Twitter, or Skype, teens have a variety of mobile services to use for conversations or gossip with friends. Messaging is often used to touch base, stay connected, or fill time when bored.

ask.fm is an application that has received a large amount of media attention for its presumed role in a number of teen suicides. Since ask.fm supports posting anonymous comments or questions to the specific social network groups you belong to, this network-based anonymity has appeared to support more blunt postings. Many of the teens in our study used this app and several discussed its use for bullying or making fun of classmates. Often, hearing about one of these events is what brings teens to use the platforms in the first place. P6 told us that she "*downloaded it because one time my friend had like drama on it so I thought ok, I'll go look at it. So I made [an account] to just see what was going on and stuff.*" She talked about people posting things like "*What about your brother? So ugly.*" and softball teammates posting about a fellow teammate who had an illness and wore a wig saying: "*Your hair is ugly, and why do you wear a wig?*"

Teens reported often trying to find out who was behind especially mean questions, sometimes using Twitter to say things like "*whoever is posting that, it's not necessary*" or asking friends if they knew who posted a particular anonymous question. They would also use the style of wording in particular posts to try to identify the poster.

Sometimes though, ask.fm was used in a fun way. P5 told us about people asking "*totally random questions and they come up with witty answers which is why I like reading them*." P6 reported how ask.fm could be used for good, in an example where a post of hers on Twitter that sounded "*like sad or whatever*" led to a question to her on ask.fm asking if she was ok. In this instance, ask.fm provided a forum to show that someone cared, without being named.

Teens reported checking social apps quite regularly throughout the day, so that they didn't "*miss anything*" that was going on with their friends. Often, checking apps would be the first task of the day. P9 told us that she checks "*Instagram and on to Twitter right when I wake up.*" P5 also checks Instagram and ask.fm "*really fast and then I get ready for school*" while P12 wants to "*see if I missed anything cool*" while sleeping.

Many of our participants reported using some kind of chat application with friends for hours each evening after school. P15 talked about joining group Skype calls: "*Everyone gets home around five and you can see when people are on so you just call the group and whoever picks up.*" P12 enjoyed using the video feature on FaceTime with friends because "*it just feels like we're together … It just feels like you're with them so sometimes you don't know what to say but we just talk about school or everything and soccer practice*." Skype and Facetime were used to facilitate continuous co-presence with others. This was done with friends that participants have seen throughout the day at school, similar to how phone calls were used in previous decades to continue conversations into the after-school hours from home.

Teens also reported using social media to follow celebrities and to post to their friends when they learned new celebrity gossip. While much of this information is reported in traditional news outlets, teens turned to different apps to learn the latest happenings from celebrities that they wanted to follow. P13 was an avid fan of the boyband One Direction and followed them on Twitter and Instagram. She also watched the entertainment show X-Factor and liked following the contestants on Twitter because they were more normal people auditioning on the show who were "*not super famous yet*." Other participants had custom apps for particular celebrities that they used to keep up to date. Teens tried to assess whether the celebrities themselves were posting or reading these sites and that seemed to influence their choices about staying on these sites and continuing to follow the celebrities. This sense of celebrities posting information themselves or answering questions directly created a much stronger perceived sense of connection that was not previously available and motivated teens to continue following or reading posts.

### *Phones for Schoolwork*

American high school students have a lot of homework, often many hours per night and typically are assigned 17.5 hours of homework per week (Burden, 2014). It is no surprise that their smartphones are often used to help with this task. Several participants would download textbooks onto a smartphone or tablet so that they did not have to bring books home. P13 discussed taking pictures of textbook pages: "*I didn't want to bring the whole book with me, but if I just take a pictures of the pages then I can have it on my phone so I don't have to bring this textbook.*" Several participants also reported using their smartphone to take notes or using the calculator app to help with homework.

Participants reported going to online sources to find information for assignments, generally through web search. P13 talked about using the smartphone during class to research a project: "*Research helps you start your essay or start doing this work. If you don't have [a phone] you have half an hour and it's going to waste.*"

Beyond web-based resources, teens often turned to friends for help with homework. When P2 is stuck on a homework problem, he will "*text friends or talk with friends and we figure stuff out together.*" P13 uses video chat to talk to friends about homework: "*If I'm trying to talk about something, I can point to it.*"

Finally, smartphones were often used to access official school communication. Several participants reported using official school Facebook groups to receive information about school activities and projects or to get homework assignments and help. For these students, Facebook was an administrative tool, and quite distinct from a social network to talk to friends. P13 used Facebook groups both for her Chemistry class and for her color guard troupe. For classwork, it was used to "*post the question like general things like, 'How is this [lab report] supposed to be organized?'*" The color guard group was used to ask questions about borrowing clothing that was needed for certain performances and verifying the times of practices.

## Discussion and Implications

Through our quantitative log analysis of all apps used, detailed voicemail diaries, and interview data, we have a better understanding of how mobile applications and services fit into teens' daily lives and why particular applications were chosen for specific uses. Some of these findings serve to confirm earlier studies that even the newer smartphones are still primarily used for various means of communication (Lenhart et al., 2010). However, we have demonstrated deeper and novel insights around how the affordances of particular apps drive teens' choices for particular types of communication. Additionally, this is one of the few studies to quantitatively log and report on actual smartphone app use (with precision beyond what self-reports and survey collect) and then blend this data with qualitative findings. The focus on teens, aged 13-18, is a novel demographic to have explored using these methods and we have highlighted significant differences between their use and previous studies of the general population. This data sets up many implications for the design of new communications applications that are targeted at younger generations.

Most notably, from the log data we have seen that the teens in our study use their mobile phones much more than what has been seen in previous work, specifically the study of Böhmer et al. (2011). While Böhmer et al. saw an average of 59 minutes per day of smartphone use, we observed 173 minutes, close to three hours of daily app usage! The teens in our study were clearly heavier smartphone users that the general public was just two years earlier. In addition we found that individual app usage duration is shorter and more bursty compared to Böhmer et al's general population (63 vs. 72 seconds on average respectively). The role of communications apps clearly is one area where teens spent a large amount of their time, accounting for two thirds of their overall smartphone use.

As teens often have varying amounts of time to kill between various activities and pick ups and drop offs, we have seen the role of the mobile device as a plastic technology (Rattenbury et al., 2008) that seeks to fill their spare time. Whether it is watching Netflix or listing to music in the car, playing games while waiting for friends to ride a roller coaster or while walking home from school, we have seen how the mobile device can act to fill the down time that the teens in our study experienced.

The desired persistence (or lack of persistent) of communication content causes significant behavior differences and choices in app use. The teens in our study want and use *both* fast-and-fleeting messages that auto-delete (e.g. Snapchat) and more formal posting of persistent content (e.g. Facebook, Twitter, Email), and we see them spending over 8 minutes per day in ephemeral apps such as Snapchat. We observed teens more carefully curate, edit, and select material when they know it will last or be shared beyond their close and trusted friends and they are very aware of managing their identity or image in this situation. They often collaboratively determine what content is shared. They also pick content they believe will generate more likes or comments and sometimes delete posts due to lack of responses from others. They tend to categorize particular apps (e.g., Facebook, Twitter, or email) as being for content that is more "formal" or official. While they realize this content "stays around forever" and others may look at their past content, teens themselves do not often re-visit their past content. The teens we observed very much live in the now and focus on not missing out on sharing immediately occurring events. This suggests that future mobile apps or services that support teen communication should make clear distinctions between material that will persist and more spontaneous temporary content. Furthermore, it raises some interesting questions around the value of historical "timeline" data views for teens. Given the value teens placed upon content "likes", re-posts or other indicators of support (and their removal of less-noticed content), this suggests interesting opportunities for tools that facilitate such monitoring and removal or archiving of this less-noticed content.

A significant portion of teens' content sharing and communication is ephemeral, consisting of temporary sharing with small groups of their closest friends, not posting in a more broadcast-like manner; there is a clear distinction between the act of *posting* content to a subgroup versus *sending a message* to a subgroup. For the teens we observed, these are very different apps where they manage very different presentations of themselves (in a classic Goffman-style (1959) presentation of self). Teens felt more comfortable to be themselves, and to share unedited quick photos of funny faces, food, or other scenes of everyday life via Snapchat for just this reason. Snapchat is not just about the photos (which are sometimes irrelevant) it is more importantly about the dynamic and temporary nature of the messages and text annotations that better supported this spontaneous informal communication. This points to the need for more fine-grained control over both reach of specific content and the duration for which it can be accessed and would give teens more room to experiment with their identities without being forever "judged" on their digital interactions. One example of this might include systems that automatically "fade away" content over time (unless that content is re-visited) so older content is removed automatically. It would also be interesting to understand how different persistence times might be used; at present content is either deleted upon being read or persists long term. It is unclear how content that lasts, for example, a day, overnight, or until the end of an event might be used. These are potential areas to explore in future apps.

The teens in our study did not "manage" their groups, circles or collections of contacts. They rarely unfriended or unfollowed. We were anticipating that teens might discuss concerns about avoiding "unfriending" because of the detrimental social message this sent (and this may still be true), however, this didn't come up in their discussions about current practices. They simply migrated to a new app or created new accounts where they formed a new social group, leaving the old group behind and they seemed to expect this approach broadly from others. This may alleviate the stigma of "unfriending" and suggests interesting new options for fading out older groups or creating groups with an expectation at the outset that it is a temporary collection that exists only while it is actively used and otherwise naturally expires. Current apps and services do not support this notion of *naturally* expiring groups or fading away over time (without deliberate, and often awkward, explicit user intervention). Future mobile apps could additionally facilitate creating deliberately temporary groups, for example around organizing for an event like a party or football game, where everyone expects the group membership to end when the event does (though content might still be archived). This creation of deliberately temporary groups and temporary user accounts would be a new opportunity/feature for tech companies who presume (until now) that both groups and accounts are long-term associations with a person.

The teens in our study were more active smartphone users (on almost every dimension) than reports indicate for adults, even with their very structured schedules of school and after-school activities. They use smartphones for longer, launch more apps, try more apps, and communicate more frequently. They are early adopters and rapid experimenters. They quickly pick up new apps and stop using others. It is difficult to determine how much of the rapid migration to new apps is motivated by this need to re-formulate new ad-hoc groups (leaving others behind with the convenient explanation that people are using a different app now), or how much was motivated by new functionality offered by the new app (e.g., auto-delete). In some cases, it may simply be wanting to try the newest, latest fad. It does suggest re-thinking strategies for user retention and new product launches that differ from tactics used for adult or broader population.

## Conclusion

Understanding how teens use their smartphones in daily life is critical to designing new mobile services and applications and is important in understanding how technology impacts identity, communication, and friendship ties. These practices evolve rapidly as new technology and apps appear, more so for teens over adults, and thus studies from even a few years ago no longer reflect current smartphone use practices. We completed a 2-week study capturing quantitative data about the daily use of mobile applications by a set of 14 teens from the San Francisco Bay area as well as detailed qualitative data about that use from voicemail diaries and interviews. To the best of our knowledge, this is the first study to quantitatively capture the application use of teenagers directly from smartphones, as well as the first to pair this data with in-depth qualitative data from interviews and diaries. Through unpacking practices around smartphone use we have shown that two-thirds of phone interactions involved communications applications and we have explored the rich details of how and why teens chose specific apps to communicate in particular scenarios. This study suggests implications for mobile communication apps that need to support nuanced practices around groups and a strong desire to manage the persistence of content with a particular emphasis on ephemeral communications. New communications experiences need to fit the complex desires of teens to manage different social faces for different groups of people.